# Capturing vehicular headway using low-cost LIDAR and processing through ARIMA prediction modeling

Azhagan Avr (Corresponding Author)
Department of Electrical and Computer Engineering,
North Carolina State University, Raleigh, NC 27695
Email: aavr@ncsu.edu

Shams Tanvir
Institute for Transportation Research and Education (ITRE),
North Carolina State University, Raleigh, NC 27695
Email: stanvir@ncsu.edu

Nagui M. Rouphail
Department of Civil, Construction, and Environmental Engineering,
North Carolina State University, Raleigh, NC 27695
Email: rouphail@ncsu.edu

Rachana Gupta
Department of Electrical and Computer Engineering,
North Carolina State University, Raleigh, NC 27695
Email: ragupta@ncsu.edu

# ABSTRACT

The paper proposes a low-cost system to capture spatial vehicle headway data and process the raw data by filtering outliers using a novel filtering process. Multiple sensors and modules are integrated to form the system. The sensors used are compact, lightweight, low-cost and have low power consumption. A single beam 1-Dimensional Light Detection and Ranging (LIDAR) was used for capturing the space headway data, a Global Positioning System (GPS) to map each data point with a timestamp and position and also a camera to capture video data with an overlay of date, time, distance and speed in real-time. The filtering technique utilizes the Autoregressive Integrated Moving Average (ARIMA) prediction modeling and mean-filtering. The data captured is stored in a Raspberry Pi module. The data is later processed by using the filtering technique to obtain the least outliers. The overall system has enabled to capture spatial headway data and speed of the vehicle at a very low cost and the data obtained can be used for car-following model analysis and speed-density analysis.

# INTRODUCTION

The motivation for this project is to develop a low-cost, dynamic and accurate prototype that would capture spatial headway data, unlike the traditional stationary devices that capture headway information from stationary positions.

In the 1970s ropes were used to measure the spatial headway between cars. But, nowadays with advances in technologies, spatial are collected through a variety of state-of-the-art techniques. Researchers have used video data to identify the headway distance (*15*) between vehicles. Some researchers have used Global Positioning System (GPS) data for speed capturing but also for estimating the headway using positional data (*16*). But in the above-mentioned techniques, the spatial headway data is being estimated and not captured, hence there is always scope for inaccuracy in such techniques.

Hence an ideal solution would be to collect the headway data using sensors like Light Detection and Ranging (LIDAR) or Radio Detection and Ranging (RADAR). At present almost all new vehicles are attached with sensors, like LIDAR or RADAR for Advanced Driver Assistance Systems (ADAS). But the data from these sensors are not accessible to users. Hence this paper would be useful to studies that use vehicles to collect spatial headway data like car-following data.

This paper discusses a cost-effective method to collect spatial headway using LIDAR, and a method to process and filter the outliers from the raw data. A 2-stage process which involves ARIMA prediction modeling in stage 1 and means filtering in stage 2 to filter out the noise in the raw data from the sensor. When the manufacturers make the raw sensor data available to users through the On-Board Diagnostics (OBD) port, the same method of noise removal can be implemented, but due to its unavailability, an external sensor is needed.

A brief on some existing techniques used to collect spatial headway data is discussed in the following section. Following the literature on existing systems, the system architecture and data collection sections explain the process of setting up the system and collecting the data. Following these sections, the method used for data processing and its algorithm is explained. Error analysis compares the chosen ARIMA model used for filtering the noise with other commonly used models for time series forecasting. In the penultimate section, the stagewise results of the filtering process are discussed and explained and finally, the study found through this paper is briefly summarized in the conclusion.

# LITERATURE REVIEW

As mentioned in the introduction, various techniques have been used to estimate spatial headway data. An image processing tool used to classify vehicles in a mixed traffic scenario is used to estimate the spatial headway (*15*) between vehicles. But the disadvantage of using cameras for headway measurement is that the camera needs to be calibrated for every location the camera is being installed, the following study shows a method that can be used to calibrate the camera (*17*), which is an added challenge. Apart from calibration, there are other challenges faced when stationary sensors are used like restrictions on installations, environmental factors and cost of installations and reinstallations (*18*). Hence dynamic data-driven methods are the way to move forward (*19*).

An alternative solution to stationary sensors could be using GPS devices to estimate the spatial headway data. A research proposes that positional and speed data obtained from GPS traces are cost-effective and accurate (*20*). Using the positional data of two known vehicles the headway is estimated (*16*). This method of dynamic headway data collection brings in the factor of dependency on the data from the leading vehicle which rules out collecting data from different commercial users independently, as at any given time 2 vehicles are needed for data collection.

In the methods mentioned above the spatial headway, are not captured directly but they are estimated. Hence the methodology used in this paper involves a naturalistic approach of data collection that is independent of other vehicles for the data.

RADAR and LIDAR both have a similar functionality except that LIDAR uses light waves and RADAR uses radio waves to measure distance. Both these devices have their set of advantages and disadvantages, but both the sensors are widely used in the latest vehicles (*7*) for ADAS. This paper uses a LIDAR device for measuring the distance.

Two different types of LIDAR such as 1-Dimensional (1D) and 2-Dimensional (2D) LIDAR were tested. The 3-Dimensional (3D) LIDAR was not tested as mapping the environment into a 3D plane was not the scope of this research. The 2D sensor can map an entire 2D plane by rotating 360 degrees, but the major drawback is that the 2D sensor has a very low sample rate because the object that the sensor was tracking has moved significantly before the sensor completes the rotation and the accuracy decreases with the object in motion, but the 2D sensor works best for mapping a stationary 2D plane. Due to the lack of compatibility of the 2D sensor with the intended application, the 1D sensors were considered. 1D sensors do not map the environment, nor rotate 360 degrees, but they produce singular light beam at a frequency between 500 – 1000 Hz which is then bounced of the surface of the leading vehicle and the reflected beams are captured and distance is measured.

They are less expensive compared to the other LIDAR devices, the LIDAR used here is priced at $125 and the device is small, low power and compact.

ARIMA is a prediction model that is used for time series forecasting (*10*). The model can predict a value based on the selected range of values. ARIMA is used to predict the spatial distance between cars. The mean filter also known as an averaging filter is used to filter the data that is rejected by the ARIMA prediction.

To summarize the literature discussed, techniques used to estimate the headway information from videos involves a tedious process of accurate calibration of the camera at every installation. GPS traces used to estimate the headway data is dependent on the GPS traces of other vehicles to estimate the headway between vehicles. But this paper uses a LIDAR that is low-cost and capable of capturing headway data at a high frequency and accuracy, also processes to filter the noise in the raw data is explained in detail. and the following questions are answered through this paper; a). Can a low-cost LIDAR capture accurate reading while in motion? b). Is the chosen ARIMA model the most appropriate to the application? and c) Are the results of the 2-stage filtering process promising?

# SYSTEM ARCHITECTURE

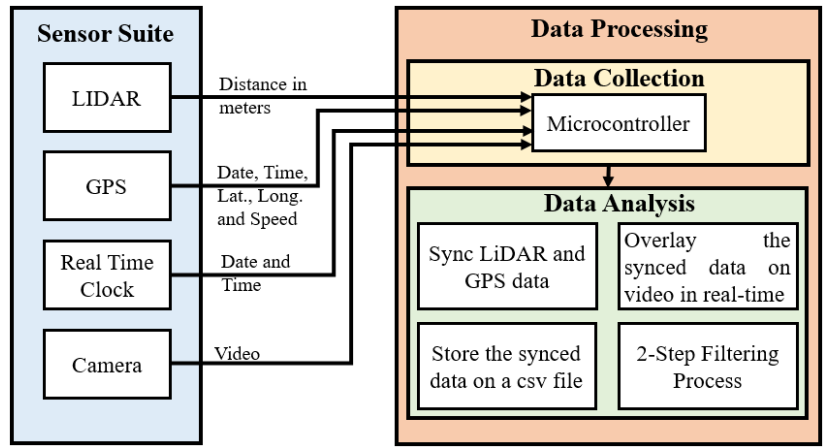

**Figure 1 System Architecture**

In the system architecture, the sensor suite consists of the LIDAR, GPS, real-time clock and a camera. The LIDAR is used for distance measurement, GPS is used to collect timestamps, speed, and positional data. The real-time clock is for updating the system time in the Raspberry pi and the video data from the camera is used as a reference to verify the data. The data output from the different devices is sent to the Data Processing unit which is a combination of an Arduino and Raspberry Pi. The Arduino receives the data captured from the GPS and LIDAR modules, organizes and syncs them together. The LIDAR has a range of 40 m and the output from the LIDAR module is in the form of pulses, whose width is converted to distance. The pulse width is expressed in the unit **μs, 10 μs = 1 cm**. The distance of the object in meters is,

$$Distance, d = \frac{Pulse\ width}{1000}\ m \quad \text{................................................................................} (1)$$

The Raspberry Pi performs the overlay of the GPS and LIDAR data on the video and stores the captured data in the CSV file. The 2-step filtering process can also be performed on the Raspberry Pi, but for convenience, visualization and efficiency of processing a large amount of data a separate laptop was used. The total cost of the entire setup would round up to $300.

## DATA COLLECTION

During data collection, the LIDAR is mounted on the hood of the car using a specially made mount, which helps in the installation process. The mount consists of 2 suction pads, a shaft, and an aluminum panel with a laser pen holder.

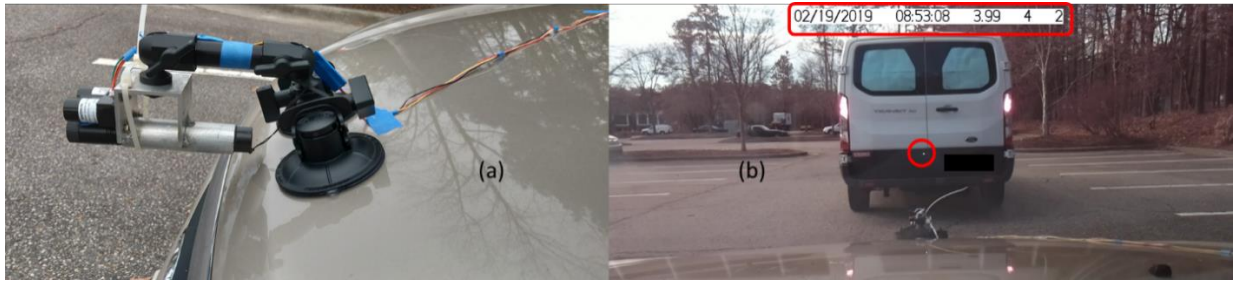

**Figure 2 (a). LIDAR and Laser pen fixed to the mount which is affixed on the hood of the car (b). The red circle highlights the green Laser light that indicates where the LIDAR is looking at.**

The 2 suction pads enable the mount to be affixed to the hood, the LIDAR is screwed to the aluminum panel and the Laser pen is slid inside the holder. There are two knobs to adjust the angle of the mount. The light waves from the LIDAR cannot be seen by the naked eye, so the Laser pen helps to make the final adjustments. As shown in Figure 2b, the image is a snapshot from a real-time video, which gives an idea about where the LIDAR is pointed (red circle).

As seen in Figure 2a, the LIDAR is connected to the in-vehicle unit through wiring, which consists of components like Arduino, Raspberry Pi, Real-time clock, GPS and Camera. The in-vehicle unit is powered through the 12V cigarette lighter port. Data is collected by following random vehicles in different traffic conditions like congested and free-flow driving situations on freeways and arterial roads.

*Video Data with Overlay:*
The other data form stored in the Raspberry Pi is the video overlaid with the date (GPS), time (GPS), spatial headway (LIDAR), follower speed (GPS), and trip id. These variables are displayed on the video for reference which is highlighted in Figure 2b.

*GPS and LIDAR Data:*
The GPS and LIDAR data captured from the respective modules are synced within the Arduino and then sent to the Raspberry Pi where data is stored in a CSV file. A sample of the data is shown below.

**Table 1: Sample dataset**

| Date | Time | Latitude | Longitude | Speed (mph) | Course Over Ground | Distance (m) | Trip Id |
|---|---|---|---|---|---|---|---|
| 2/19/2019 | 10:12:40 AM | 35.00008 | -78.6646 | 18 | 286.52 | 7.74 | 6 |
| 2/19/2019 | 10:12:40 AM | 35.00008 | -78.6646 | 18 | 286.52 | 7.79 | 6 |
| 2/19/2019 | 10:12:40 AM | 35.00008 | -78.6646 | 18 | 286.52 | 7.81 | 6 |
| 2/19/2019 | 10:12:41 AM | 35.78681 | -78.6646 | 18 | 287.73 | 0.14 | 6 |
| 2/19/2019 | 10:12:41 AM | 35.78681 | -78.6646 | 18 | 287.73 | 7.62 | 6 |
| 2/19/2019 | 10:12:41 AM | 35.78681 | -78.6646 | 18 | 287.73 | 7.78 | 6 |

The first and second column consists of the date and time, the third and fourth column consists of the vehicle's latitude and longitude, 5$^{th}$ column is the speed of the vehicle in miles per hour, the course over ground gives the direction in which the vehicle is moving and all the above-mentioned data are obtained from the GPS module. The 7$^{th}$ column is the distance data from the LIDAR module in meters. The final column is the trip-id used as an identification number to identify each trip individually.

## DATA ANALYSIS

The data stored on the CSV file are subjected to post-processing to remove noise. An algorithm was designed based on ARIMA modeling and mean filtering to remove the noise in the data, which is explained in the following sections.

### Simple Exponential Smoothing (SES):

The exponential smoothing window is used to smoothen the time series data. The past values are given equal weight in a moving average window, but in simple exponential smoothing, weights are exponentially increased over time (*8*). The exponentially increasing weights help in providing less weight to the values away from the data point of interest (*9*). Exponential smoothing is given by the equation (*8*):

$$S_t = \alpha \cdot x_t + (1 - \alpha) \cdot s_{t-1} = s_{t-1} + \alpha \cdot (x_t - s_{t-1}) \dots\dots\dots\dots\dots\dots\dots\dots\dots\dots\dots\dots\dots\dots (2)$$

where $\alpha$ is the smoothing factor with values from $0 < \alpha < 1$. $S_t$ is the weighted average of the most recent observation $x_t$ and $S_{t-1}$ is the previously smoothed data. Large values of $\alpha$ provide greater weight to recent changes in the data, while small values of $\alpha$ are less responsive to recent changes. Hence the value of $\alpha$ depends on the data and the application. In this paper, the simple exponential smoothing model is responsive to recent changes because each measured reading is independent of the previous readings i.e. the past values do not have an impact on present reading.

### Auto-Regressive Integrated Moving Average (ARIMA):

ARIMA (0,1,1) prediction modeling performs simple exponential smoothing (*11*). The model performs incremental exponential smoothing by gradually discounting the past values i.e. lesser weights are given to values away from the current observation (*12*), by which less smoothing and more response to recent changes are achieved (*8*). This prediction helps in filtering out the noise in the raw data. When the difference between the predicted value and the measured value is less than 2m (threshold 1 - TH1) then the reading is a valid data point, otherwise, the measured value is considered as a noisy reading and sent to the next stage for filtering. Noises are of three types namely, system noise that is caused by weak signals received after reflection from the vehicle, noise due to objects in the environment like trees, bushes, lamp posts and noise due to change in car-following events. etc.

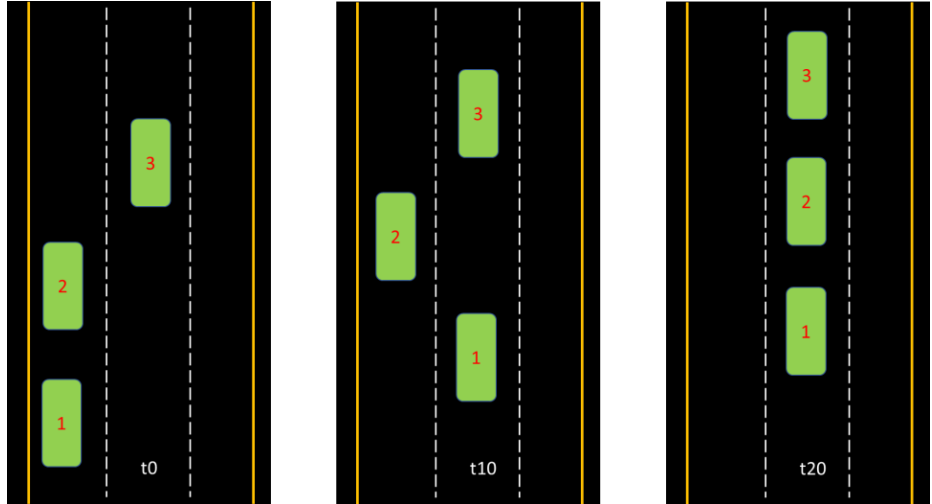

**Figure 3 illustrates the change in car-following events.**

But some valid readings resemble like noise. For example, as illustrated in Figure 3, assuming at every instant 't' the distance between any two vehicles is 5 m and the length of each vehicle is 10 m. So, at instant t0, the vehicle 1 (equipped) is following vehicle 2 at 5 m and after 10 instants of time vehicle 1 changes lane to follow vehicle 3 at 20 m approximately. So, this sudden change in value from 5 m to 20 m is called a change in the car-following event. Similarly, at instant t20 vehicle 2 moves between vehicle 1 and vehicle 3, hence now the distance measured changes from 20 m to 5 m. Hence these changes in car-following events are like the noise values with sudden upward or downward spike in readings. To avoid throwing out such values mean filtering is performed.

*Mean Filter:*

As mentioned earlier, a sudden change in spatial headway can be due to weak signals or random objects or due to change in the car-following events. These types of readings are filtered out by ARIMA prediction modeling. Hence through mean filtering the three types of readings are differentiated and the change in car-following event readings are retained and the noisy values are discarded. The method of differentiating the readings is explained below;

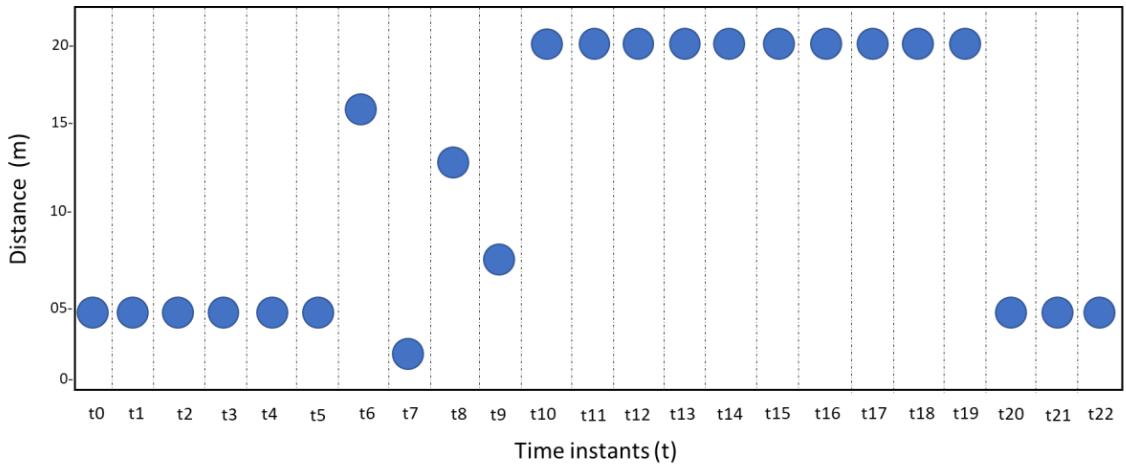

**Figure 4 illustrates the raw data from the LIDAR for the scenarios shown in Figure 3.**

With reference to Figure 3, Figure 4 illustrates the raw readings from the LIDAR module. The readings at t6, t8, and t9 are due to random objects and reading at t7 is due to a weak reflected signal from the objects. The remaining data points are reading of the vehicles. Readings captured from instant t0 to t5 are retained as valid data points by ARIMA prediction modeling. At t6 there is a sudden spike in data which is rejected by the ARIMA model, once rejected by the ARIMA model, mean filtering is performed, by averaging the values from t6 to t10. So, the actual value at t6 is 17m but the average of the values from t7 to t11 is 14.6 m, the difference between the actual value at t7 and averaged value is 12.4 m, so only when the difference between the actual value and the averaged value is less than 1m (threshold 2 as TH2) the value is retained. So, in this case, the value at t6 is discarded as a noisy value. Similarly, corresponding values from t7 to t9 are also discarded.

The next change in the event occurs from t9 to t10 as mentioned before ARIMA rejects the value of 20 m at t10 due to the sudden change in value. As performed earlier, the average from t10 to t14 is 20 m and the difference between t10 and the average is less than 1m so the sudden change in value is accepted as a valid data point. Similar processing happens in the transition from t19 to t20.

Hence through a combination of ARIMA prediction and mean filtering valid data points are retained and invalid data points are discarded. The algorithm for this process is explained in detail below.

## ALGORITHM

There are two levels of filtering. In the first level, a window size (N) of 30 values is considered and the 31st value (N+1th value which is, P) is predicted through ARIMA forecasting if the difference between P and the measured value M is less than TH1 then the reading is a valid data point. But if the difference between P and M is greater than TH1 then the value is sent to the next level of filtering. In the second stage of filtering, the measured value M is assumed to be the starting of a new car-following event or noise. So, the average of M and the next 4 readings are taken and if the difference between the M and the average is less than TH2, then the measured value M is a valid data point or else M is a noisy reading. The discarded values are not considered for populating the sampling window which is used for predicting the value P. The two threshold values TH1 and TH2, are identified through trial and error to achieve best-filtered data output.

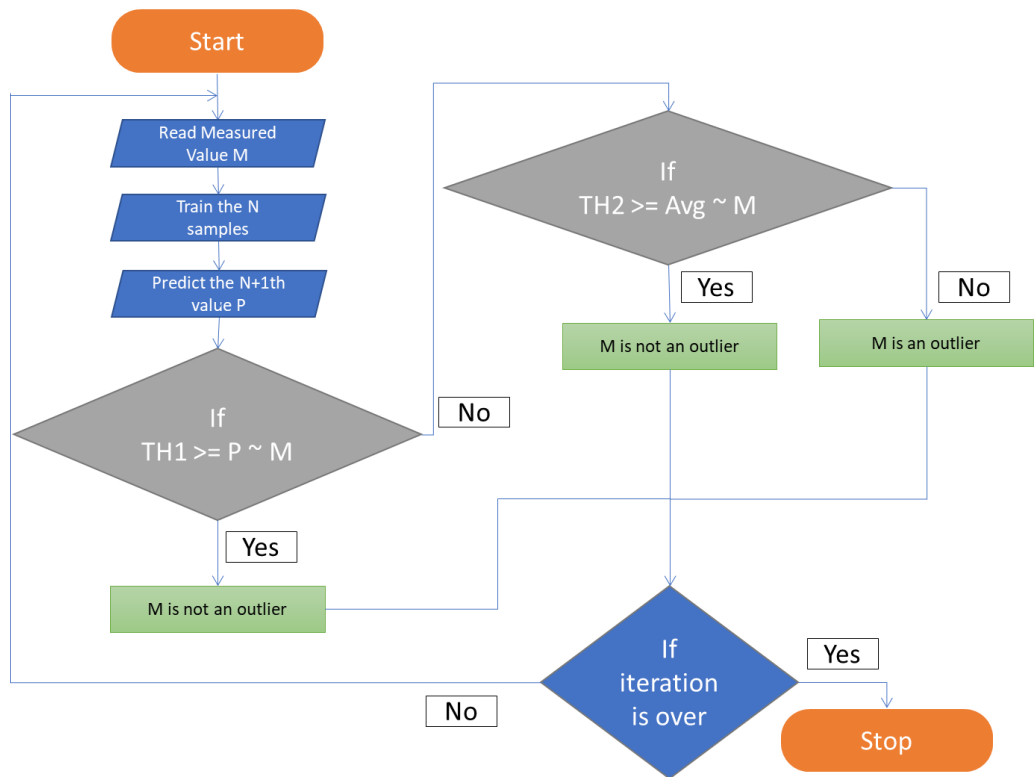

**Figure 5 Shows a Flowchart of the 2 step-filtering process**

## ERROR ANALYSIS

The main reason for choosing ARIMA (0,1,1) was because of its exponential smoothing model. Table 2 shows the error analysis between the different ARIMA models (*10*) commonly used for time series forecasting. To perform the error analysis, ground truth data were collected by physical measurements in a controlled environment.

A test vehicle parked at 100 m from the starting point, the equipped vehicle was driven towards the stationary vehicle at a constant speed of 10 mph, the LIDAR started recording from a stipulated distance of 26 m from the stationary vehicle and the equipped vehicle was stopped at 10 m before the parked vehicle. Since the equipped vehicle was driven at a constant speed, and the initial LIDAR recording distance is known, the ground truth value is calculated for every instant a LIDAR reading was available in that range.

**Table 2: Error Analysis**

| ARIMA Model | Ground Truth Vs Prediction | | | | Ground Truth Vs Filtered Data | | | |
|---|---|---|---|---|---|---|---|---|
| | MSE | RMSE | MAPE | MAE | MSE | RMSE | MAPE | MAE |
| (0,0,0) | 17.28 | 4.16 | 23.27 | 3.40 | 1.38 | 1.18 | 5.72 | 1.04 |
| (0,1,0) | 1.49 | 1.22 | 3.03 | 0.66 | 0.98 | 0.99 | 2.23 | 0.50 |
| (0,1,1) | 0.13 | 0.36 | 1.79 | 0.30 | 0.08 | 0.28 | 1.22 | 0.22 |
| (1,0,0) | 1.01 | 1.01 | 2.74 | 0.54 | 0.96 | 0.98 | 2.45 | 0.50 |
| (1,1,0) | 1.08 | 1.04 | 2.64 | 0.53 | 0.96 | 0.98 | 2.27 | 0.49 |
| (0,2,1) | 2.61 | 1.62 | 4.04 | 0.88 | 1.39 | 1.18 | 3.05 | 0.68 |

The error analysis between ground truth and predicted value proves the accuracy of the ARIMA prediction with respect to the ground truth and the error analysis between the ground truth and the filtered data (final valid data points) proves the accuracy of the sensor with respect to the ground truth.

Table 2 highlights the performance of different ARIMA models. ARIMA (0,1,1) has the least error performance when compared with the other ARIMA models. Both the predicted data (through ARIMA prediction) and the filtered data (final valid data points) are compared to the ground truth for accurate error analysis. The graph below shows how the ARIMA (0,1,1) predicted data and the filtered data fares when compared to the ground truth values.

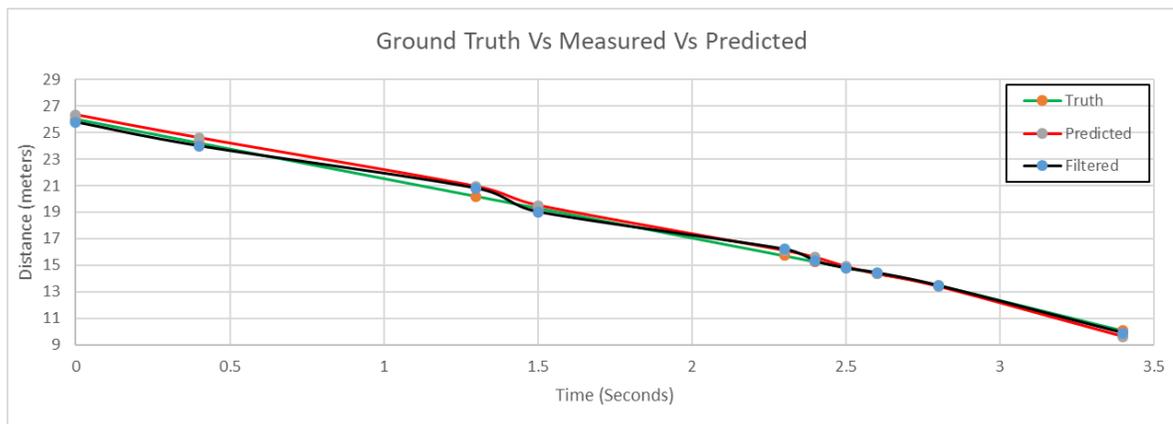

**Figure 6 compares ground truth values with predicted and filtered data.**

# RESULTS

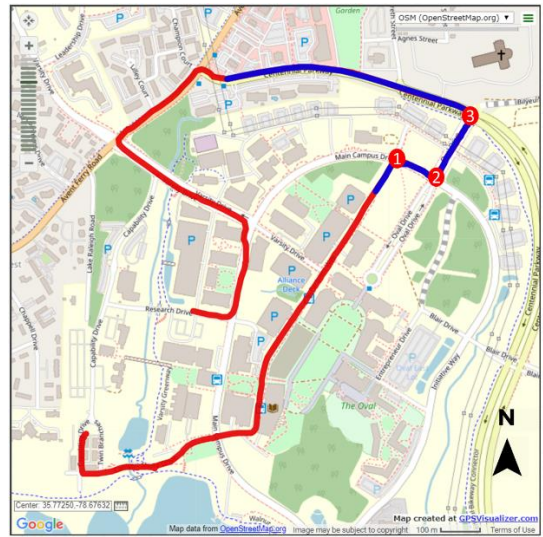

**Figure 7 Route traveled for collecting the data**

Figure 7 shows the trip map, the highlighted red route is the entire trip, the highlighted blue route is the section chosen for filtering. The three red dots are three instances of change in the car-following event. In the first instance the leading vehicle turns right eastbound on main campus drive, Raleigh, in the second event the leading vehicle turns left, northbound on Oval drive, Raleigh and in the third event the leading vehicle was turning left, westbound on Centennial Parkway, Raleigh.

The equipped vehicle was following the leading vehicle before and after the change in car-following events. The route map is generated using the GPS data in the system. Following are the stage-wise results of the filtering process, the data shown below belongs to the blue line on the map.

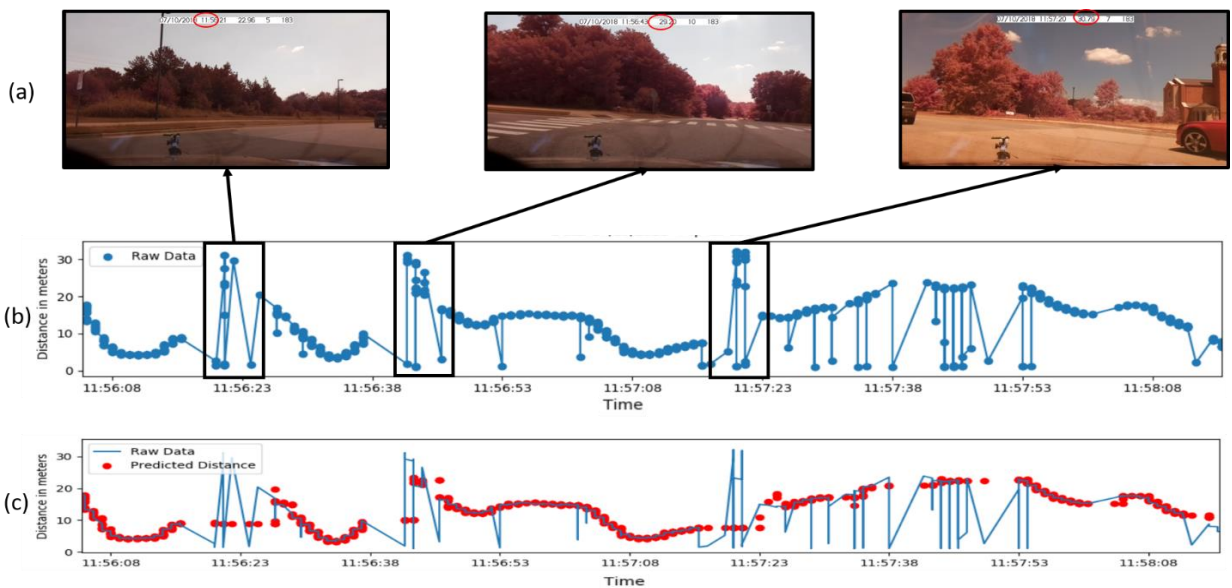

**Figure 8 (a). The three change in car-following events (b). The plot highlights the high noise sections in the data (c). Red dots denote the data predicted using ARIMA (0,1,1) modeling**

In Figure 8b the data points with noisy spikes are highlighted. The smooth curves denote the car following section and the oscillating spikes that are noisy readings which are followed by the change in car-following readings. As shown by the images in Figure 8a, values are captured even if there are no vehicles present. These noises are due to objects in the environment such as lamp posts, bushes, objects on the shoulder, etc. The two-stage filtering process helps in detecting these changes in the car-following events. Figure 8c shows the predicted data and the measured data (raw data), using these two datasets the 2-stage filtering is performed and the following result is obtained.

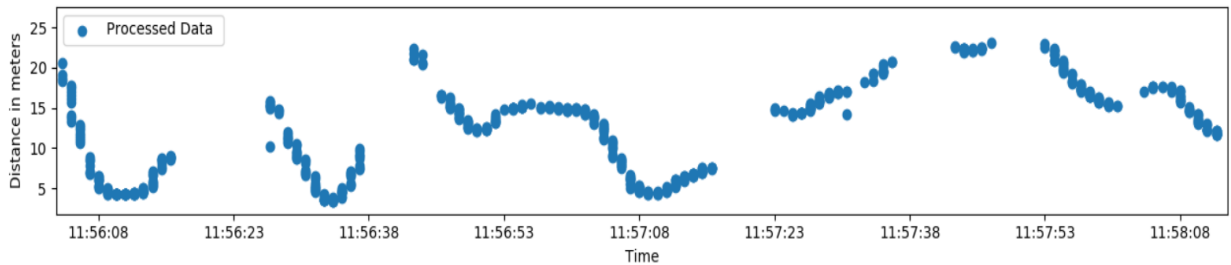

**Figure 9 The plot shows the fully processed output**

In Figure 9, the blue dots are the filtered output data and the noisy data are clearly filtered out and only the spatial headway data is obtained as the final output.

## CONCLUSION

Evolving from the stationary mode of collecting spatial headway to the dynamic mode of data collection seems to be a promising method for the future. The paper describes a system comprising of LIDAR which is used to capture vehicular headway integrated with GPS for timestamps, speed, and positional data and a camera for data validation. Also, a novel 2-stage noise removal process is explained in detail. Error analysis shows that the ARIMA (0,1,1) has the lowest Mean Square Error (MSE) of 0.13 m w.r.t the ground truth and similarly the final output after the 2-stages of filtering has an MSE of 0.08 m w.r.t the ground truth. The final output data is a series of car-following data. With autonomous vehicles to hit the market in the near future, such mode of data collection would be more efficient compared to stationary devices.

If vehicle manufacturers provide the data from LIDAR like sensors fixed in the vehicle to the users through the OBD port, then the need for an external sensor can be ruled out and the data can be directly obtained from the OBD port. However, the data might need to be processed since the raw data is being retrieved. The OBD data can be wirelessly transferred to smartphones (*21*) and can be logged on remote servers. Researchers can post-process the data for noise reduction and this data can be used to study driving behavior through macroscopic traffic stream models and microscopic car-following regimes such as Wiedemann model (*13*). This data may be used for Fuel Efficiency Score (FES) analysis to study trip-based fuel consumption measures (*14*).

## ACKNOWLEDGMENTS
The authors of this paper would like to thank Thomas chase for his valuable support throughout this project. This research is funded partially by the US Department of Energy (DOE) Advanced

**AUTHOR CONTRIBUTIONS**

The authors confirm contribution to the paper as follows: study conception and design: Azhagan A., Shams T., Nagui R.; hardware: Azhagan A., data collection: Azhagan A.; data analysis: Azhagan A., Shams T., Nagui R., Rachana G., draft manuscript preparation: Azhagan A., Shams T., Nagui R., Rachana G. All authors reviewed the results and approved the final version of the manuscript.